\documentclass[aps,prl,nofootinbib,floatfix,twocolumn,showpacs,superscriptaddress]{revtex4}

\usepackage{pstricks}
\usepackage{graphicx}
\usepackage{dcolumn}
\usepackage{bm}
\usepackage{rotating}
\usepackage{amsmath,amssymb,amsfonts}
\usepackage{pifont}

\newcommand{\hermes}{\textsc{Hermes}}
\newcommand{\compass}{\textsc{Compass}}

\newcommand{\desy}{\textsc{Desy}}
\newcommand{\hera}{\textsc{Hera}}

\newcommand{\pythia}{\textsc{Pythia6}}

\newcommand{\pperp}{P_{h\perp}}
\newcommand{\Vpt}{\ensuremath{\V{p}_{T}^{}}}

\newcommand{\VKt}{\ensuremath{\V{K}_{\!T}}}
\newcommand{\VSt}{\ensuremath{\V{S}_{T}}}

\newcommand{\sivers}{\ensuremath{f_{1{T}}^{\perp}}}
\def\siversf#1{\ensuremath{f_{1{T}}^{\perp #1}}}

\def\V#1{\ensuremath{\boldsymbol{#1}}}

\begin{document}


\title{Observation of the Naive-T-odd Sivers Effect in Deep-Inelastic Scattering}

\date{\bf \today}


\def\groupargonne{\affiliation{Physics Division, Argonne National Laboratory, Argonne, Illinois 60439-4843, USA}}
\def\groupbari{\affiliation{Istituto Nazionale di Fisica Nucleare, Sezione di Bari, 70124 Bari, Italy}}
\def\groupbeijing{\affiliation{School of Physics, Peking University, Beijing 100871, China}}
\def\groupcolorado{\affiliation{Nuclear Physics Laboratory, University of Colorado, Boulder, Colorado 80309-0390, USA}}
\def\groupdesy{\affiliation{DESY, 22603 Hamburg, Germany}}
\def\groupzeuthen{\affiliation{DESY, 15738 Zeuthen, Germany}}
\def\groupdubna{\affiliation{Joint Institute for Nuclear Research, 141980 Dubna, Russia}}
\def\grouperlangen{\affiliation{Physikalisches Institut, Universit\"at Erlangen-N\"urnberg, 91058 Erlangen, Germany}}
\def\groupferrara{\affiliation{Istituto Nazionale di Fisica Nucleare, Sezione di Ferrara and Dipartimento di Fisica, Universit\`a di Ferrara, 44100 Ferrara, Italy}}
\def\groupfrascati{\affiliation{Istituto Nazionale di Fisica Nucleare, Laboratori Nazionali di Frascati, 00044 Frascati, Italy}}
\def\groupgent{\affiliation{Department of Subatomic and Radiation Physics, University of Gent, 9000 Gent, Belgium}}
\def\groupgiessen{\affiliation{Physikalisches Institut, Universit\"at Gie{\ss}en, 35392 Gie{\ss}en, Germany}}
\def\groupglasgow{\affiliation{Department of Physics and Astronomy, University of Glasgow, Glasgow G12 8QQ, United Kingdom}}
\def\groupillinois{\affiliation{Department of Physics, University of Illinois, Urbana, Illinois 61801-3080, USA}}
\def\groupmichigan{\affiliation{Randall Laboratory of Physics, University of Michigan, Ann Arbor, Michigan 48109-1040, USA }}
\def\groupmoscow{\affiliation{Lebedev Physical Institute, 117924 Moscow, Russia}}
\def\groupnikhef{\affiliation{National Institute for Subatomic Physics (Nikhef), 1009 DB Amsterdam, The Netherlands}}
\def\groupstpetersburg{\affiliation{ Petersburg Nuclear Physics Institute, Gatchina, Leningrad region 188300, Russia}}
\def\groupprotvino{\affiliation{Institute for High Energy Physics, Protvino, Moscow region 142281, Russia}}
\def\groupregensburg{\affiliation{Institut f\"ur Theoretische Physik, Universit\"at Regensburg, 93040 Regensburg, Germany}}
\def\grouprome{\affiliation{Istituto Nazionale di Fisica Nucleare, Sezione Roma 1, Gruppo Sanit\`a and Physics Laboratory, Istituto Superiore di Sanit\`a, 00161 Roma, Italy}}
\def\grouptriumf{\affiliation{TRIUMF, Vancouver, British Columbia V6T 2A3, Canada}}
\def\grouptokyo{\affiliation{Department of Physics, Tokyo Institute of Technology, Tokyo 152, Japan}}
\def\groupamsterdam{\affiliation{Department of Physics and Astronomy, Vrije Universiteit, 1081 HV Amsterdam, The Netherlands}}
\def\groupwarsaw{\affiliation{Andrzej Soltan Institute for Nuclear Studies, 00-689 Warsaw, Poland}}
\def\groupyerevan{\affiliation{Yerevan Physics Institute, 375036 Yerevan, Armenia}}
\def\groupnone{\noaffiliation}


\groupargonne
\groupbari
\groupbeijing
\groupcolorado
\groupdesy
\groupzeuthen
\groupdubna
\grouperlangen
\groupferrara
\groupfrascati
\groupgent
\groupgiessen
\groupglasgow
\groupillinois
\groupmichigan
\groupmoscow
\groupnikhef
\groupstpetersburg
\groupprotvino
\groupregensburg
\grouprome
\grouptriumf
\grouptokyo
\groupamsterdam
\groupwarsaw
\groupyerevan


\author{A.~Airapetian} \groupgiessen \groupmichigan
\author{N.~Akopov}  \groupyerevan
\author{Z.~Akopov}  \groupyerevan
\author{E.C.~Aschenauer}  \groupzeuthen
\author{W.~Augustyniak}  \groupwarsaw
\author{A.~Avetissian}  \groupyerevan
\author{E.~Avetisyan}  \groupdesy
\author{A.~Bacchetta}  \groupdesy
\author{B.~Ball}  \groupmichigan
\author{N.~Bianchi}  \groupfrascati
% ...

\collaboration{The \hermes\ Collaboration} \noaffiliation

\begin{abstract}
Azimuthal single-spin asymmetries of lepto-produced pions and charged kaons 
were measured on a transversely polarized hydrogen target. 
Evidence for a naive-T-odd, transverse-momentum-dependent parton distribution function
is deduced from non-vanishing Sivers effects for \( \pi^+ \), \(\pi^0\), and
\( K^\pm \),  as well as  in the difference of  the \( \pi^+\) and \(\pi^- \) cross sections.
\end{abstract}

\pacs{13.60.-r, 13.88.+e, 14.20.Dh, 14.65.-q}

\maketitle

The ongoing experimental effort in spin-dependent high-energy scattering 
and attendant theoretical work continue to indicate that the spins of the quarks and 
gluons are not sufficient to explain the nucleon spin~\cite{deFlorian:2008mr}. The 
investigation of the only remaining contribution, that of orbital angular momentum 
of the constituents, is clearly essential.
Transverse-momentum-dependent parton distribution functions are recognized 
as a tool to study spin-orbit correlations, hence providing experimental 
observables  for  studying orbital angular momentum. 
One particular example is the Sivers function \siversf{}~\cite{Sivers:1990cc},
describing the correlation between the momentum direction of
the struck quark and the spin of its parent nucleon. This correlation is commonly 
defined as the Sivers effect.
A non-vanishing \siversf{} contributes to, e.g., single-spin asymmetries (SSAs)
in semi-inclusive deep-inelastic scattering (DIS) off transversely polarized protons, 
\(ep^\uparrow \rightarrow e'hX\), where \(h\) is a hadron detected in 
coincidence with the scattered lepton \(e'\).

For a long time, 
transverse 
SSAs had been assumed to be negligible in
hard scattering processes. They are odd under naive time reversal, i.e., 
time reversal of three-momenta and angular momenta,
and thus require interference of amplitudes with different helicities and phases. 
In QED and perturbative QCD, these ingredients are 
suppressed~\cite{PhysRev.143.1310,Kane:1978nd}.
Therefore, in semi-inclusive DIS they must 
be ascribed to the non-perturbative parts in the cross section, i.e., to 
specific parton distribution
and fragmentation functions, commonly categorized as being naive-T-odd.
The idea of a naive-T-odd quark distribution function goes back to an 
interpretation~\cite{Sivers:1990cc} of large left-right asymmetries observed in 
pion production in the collision of unpolarized 
with transversely polarized nucleons~\cite{Antille:1980th}.
It was argued that such asymmetries could be attributed to a  
left-right asymmetry in the distribution of unpolarized quarks in
transversely polarized nucleons, i.e., an asymmetry that exists before
the pion is formed in the fragmentation process, and that does not vanish at 
high energies. 
A decade after an initial proof~\cite{Collins:1993kk} that this
distribution function, now termed the Sivers function, must
vanish because of time-reversal invariance of QCD, 
it was realized through the pioneering work in Ref.~\cite{Brodsky:2002cx} and
subsequently in Refs.~\cite{Collins:2002kn,Ji:2002aa,Belitsky:2002sm} 
that this proof applies only to transverse-momentum-integrated  
distribution functions. A gauge link, previously neglected
in the definition of gauge-invariant distribution functions,
invalidates the original proof for the case of transverse-momentum-dependent
distribution functions. The gauge link provides the 
phase for the interference (required for naive-T-oddness),
and can be interpreted as an interaction of the struck 
quark with the color field of the target remnant~\cite{Burkardt:2003yg}.

The inclusion of the gauge link has profound consequences on
factorization proofs and on the concept of universality, which are of
fundamental relevance for high-energy hadronic physics. 
A direct QCD prediction is a Sivers effect in the Drell--Yan process that 
has the opposite sign compared to the one in semi-inclusive DIS~\cite{Collins:2002kn}.
For hadron production in proton-proton collisions 
the situation is more intricate~\cite{Bacchetta:2005rm},
leading to a violation of standard factorization
and universality, even for the case of unpolarized 
collisions~\cite{Collins:2007nk}.
Therefore, the study of the Sivers effect in semi-inclusive DIS and other 
processes is of utmost importance for our understanding of 
high-energy scattering involving hadrons.

The Sivers effect has been related to the orbital motion of quarks inside a 
transversely polarized nucleon since the seminal work in Ref.~\cite{Sivers:1990cc}.
In the calculation of Ref.~\cite{Brodsky:2002cx}, it became clear that orbital angular
momentum of quarks is needed for a non-vanishing Sivers effect as it arises
through overlap integrals of wave-function components with different orbital
angular momenta. However, no quantitative relation has yet
been found between \sivers\ and the orbital angular
momentum of quarks. One faces a similar 
quandary with the anomalous magnetic moment \(\kappa\) of the nucleon: 
it also requires wave function components with non-vanishing quark orbital 
angular momentum without constraining the {\em net} orbital angular 
momentum~\cite{Burkardt:2005km}. 
Indeed, \sivers\  involves overlap integrals between the same 
wave function components that  also appear in the expressions for 
\(\kappa\) as well as for the total angular momentum in 
the Ji relation~\cite{Ji:1997ek} for the nucleon-spin 
decomposition~\cite{Brodsky:2002cx,Burkardt:2005km}.

An interesting link between \(\kappa\) and \siversf{}
was suggested in Ref.~\cite{Burkardt:2002ks}: the
sign of the quark-flavor contribution to \(\kappa\)
determines the sign of \siversf{}\
for that quark flavor. If the final-state interactions 
are attractive, as one would assume for the
confining color force, a positive flavor contribution to \(\kappa\)
leads to a negative
\siversf{}. (The sign and angle definitions follow the {\it Trento Conventions}~\cite{Bacchetta:2004jz}.)

In semi-inclusive DIS, \sivers\ leads to SSAs in the distribution 
of hadrons in the azimuthal angle about the virtual-photon direction.
In general, azimuthal SSAs provide important information not only about the 
Sivers function but also about other distribution and fragmentation functions.  
For example, transversity~\cite{Ralston:1979ys}, 
describing the distribution of transversely polarized 
quarks in
transversely polarized nucleons, 
combined with the naive-T-odd
Collins fragmentation function~\cite{Collins:1993kk}, also leads  to SSAs. 
The keys to extracting different combinations of the various distribution 
and fragmentation functions are their different dependences on the two 
azimuthal angles \(\phi\) and \(\phi_S\) 
of the hadron momentum $\boldsymbol{P}_h$ 
and of the transverse component  \VSt\ of the target-proton spin, 
respectively, about the virtual-photon direction (cf.~\cite{Bacchetta:2004jz}).
The Sivers effect manifests itself as a
\(\sin(\phi-\phi_S)\) modulation in the azimuthal distribution~\cite{Boer:1998nt}.

In this Letter clear evidence for a non-vanishing Sivers function is reported. 
The \(\sin(\phi-\phi_S)\) modulations in semi-inclusive DIS are measured for 
pions and charged kaons, as well as in the 
difference between the $\pi^+$ and $\pi^-$ cross sections, providing
sensitivity to \sivers\ for both valence and sea quarks.

The data reported here were recorded during the 2002--2005 running period of
the \hermes\ experiment using a  transversely nuclear-polarized hydrogen 
gas target internal to the $27.6$\,GeV  \hera\ lepton (\(e^+\) or \(e^-\))
storage ring at \desy. 
The open-ended target cell was fed by an atomic-beam 
source~\cite{Nass:2003mk} based on Stern--Gerlach separation 
combined with radio-frequency transitions of hyperfine states.
The nuclear spin direction was
flipped at 1--3\,min time intervals, while both nuclear polarization and
the atomic fraction inside the target cell were continuously
measured~\cite{hermes:BRP_TGA}. 
The average magnitude of the proton-polarization component 
perpendicular to the 
lepton-beam direction was $0.725\pm 0.053$. 

Scattered leptons and coincident hadrons were detected by
the \hermes\ spectrometer~\cite{Ackerstaff:1998av}.
Leptons were identified with an efficiency exceeding 98\% and a
hadron contamination of less than 1\%. Charged hadrons with
momentum \(2\,\text{GeV} 
< |\boldsymbol{P}_h| < 15\,\text{GeV}\)  
were identified using a dual-radiator ring-imaging {\v C}erenkov detector 
(RICH)~\cite{Akopov:2000qi}. 
For this a hadron-identification algorithm was employed
that takes into account the topology of the whole event,
in contrast to the track-level algorithm in previous analyses~\cite{Airapetian:2004tw}.
Events were selected subject to the 
requirements 
$Q^2>1$\,GeV$^2$, $W^2 >10$\,GeV$^2$, $0.1 < y <  0.95$, and $0.023<x< 0.4$, where
\(Q^2\equiv -q^2\equiv -(k-k')^2\), \(W^2\equiv(P+q)^2\),
\(y\equiv (P\cdot q)/(P\cdot k)\),  and  
\(x\equiv Q^2/(2P\cdot q)\).
Here, \(P\), \(k\), and \(k'\) represent
the four-momenta of the target proton, the incident lepton, and the scattered lepton,
respectively. Coincident hadrons were accepted if \(0.2<z<0.7\), 
where $z\equiv(P\cdot P_h)/(P\cdot q)$.

The cross section for semi-inclusive production of hadrons 
using an unpolarized lepton beam on a transversely polarized target can be 
written as~\cite{Mulders:1996dh,Boer:1998nt,Bacchetta:2006tn}
\begin{eqnarray} 
\sigma(\phi,\phi_S) \!\!
  &=& 
      \!\! \sigma_{_{UU}} \lbrace 1+ 2\langle\cos\phi\rangle_{_{UU}}\cos\phi + 2\langle\cos2\phi\rangle_{_{UU}} \cos2\phi \nonumber\\
   &+& \!\! \mid \!\! \VSt \!\! \mid  \! \left[ 2 \langle\sin(\phi\!-\!\phi_S)\rangle_{_{UT}} \sin(\phi-\phi_S)+ \ldots \right] \rbrace
, \, \label{eq:sigma_UT}
\end{eqnarray}
where five sine modulations contribute to the  polarization-dependent part, but, for convenience, 
only the \(\sin(\phi-\phi_S)\) modulation (the Sivers term), is written out explicitly.
Here, the subscript \(UT\) denotes unpolarized beam and transverse 
target polarization (with respect to the virtual-photon direction), 
while \(\sigma_{_{UU}}\) represents the \(\phi\)-independent part of the 
polarization-independent cross section.
The \(\sin(\phi-\phi_S)\) amplitude can be interpreted in the quark-parton model as~\cite{Boer:1998nt}
\begin{equation}\label{eq:QPM-sivers}
2\langle\sin{(\phi-\phi_S)}\rangle_{_{UT}} 
= -\frac{\sum_q e_q^2  \siversf{,q}(x,p_T^2) \otimes_{\mathcal W} D_1^q(z,K_T^2)}{\sum_q e_q^2 f_1^q(x,p_T^2) \otimes D_1^q(z,K_T^2)},
\end{equation}
where the sums run over the quark flavors, the \(e_q\) are the quark
charges, and \(f_1\) and \(D_1\) are the spin-independent quark
distribution and fragmentation functions, respectively. The symbol \(\otimes\) (\(\otimes_{\mathcal W}\))
represents a (weighted) convolution integral over intrinsic and fragmentation transverse momenta
\Vpt\ and \VKt, respectively.

The amplitudes of the five sine modulations in Eq.~\eqref{eq:sigma_UT} were extracted
simultaneously to avoid cross contamination.
For this a maximum-likelihood fit was used~\cite{Diefenthaler2009},
with the data alternately binned in $x$, $z$, and 
\(P_{h\perp}\equiv | \boldsymbol{P}_h - \frac{(\boldsymbol{P}_h \cdot 
\boldsymbol{q})\boldsymbol{q}}{|\boldsymbol{q}|^2}| \), 
but unbinned in $\phi$ and $\phi_S$.
A sixth term, arising from the small but non-vanishing 
target-spin component that is longitudinal to the virtual-photon
direction when the target is polarized perpendicular to the 
beam direction~\cite{Diehl:2004}, was also included in the fit.

A scale uncertainty of 7.3\%  on the extracted Sivers amplitudes arises from the 
accuracy of the target-polarization determination.
Inclusion in the fit of estimates~\cite{Giordano:2009hi}
for the \(\cos\phi\) and \(\cos2\phi\) amplitudes of the unpolarized cross section
had negligible effects on the amplitudes extracted.
Possible contributions~\cite{Diehl:2004} to the amplitudes from the non-vanishing
longitudinal target-spin component were estimated based on measurements of SSAs on 
longitudinally polarized protons~\cite{Airapetian:1999tv,Airapetian:2001eg} 
and included in the systematic uncertainty.
Effects from the hadron identification using the RICH,
the geometric acceptance, smearing due to detector resolution, 
and radiative effects are not corrected for in the data. Rather, the size of 
all these effects was estimated using a simulation tuned to the data, 
which involved a fully differential polynomial fit to the measured 
azimuthal amplitudes~\cite{Pappalardo:2008zza}. The result was included 
in the systematic uncertainty and constitutes the largest contribution.

%
%

\begin{figure}
\centering
\includegraphics[scale=0.47]{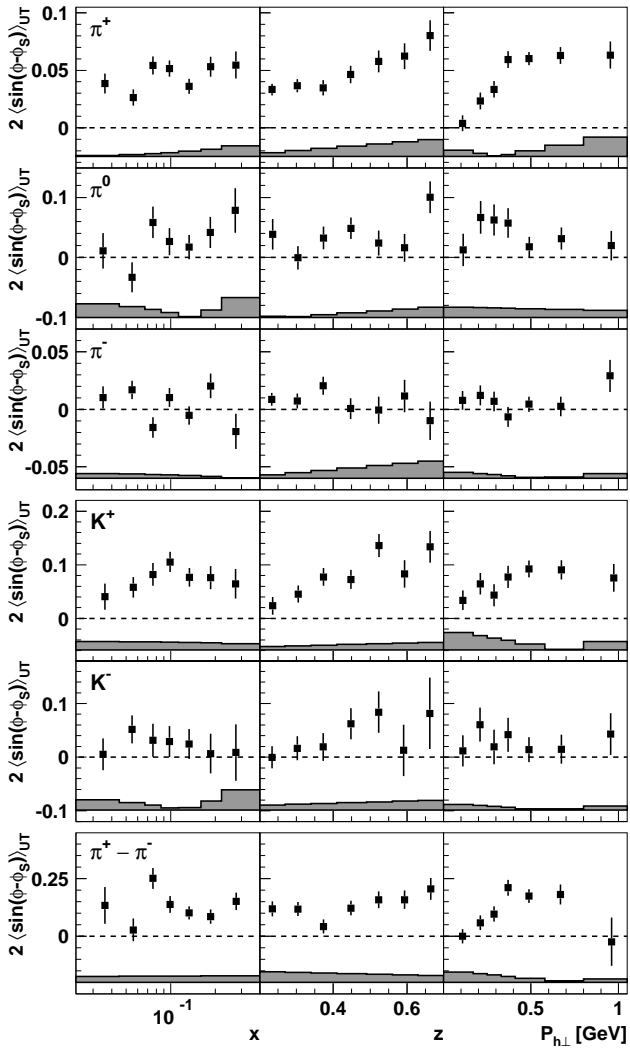}
\caption{\label{fig:main-results-all} Sivers amplitudes for pions,
  charged kaons, and the pion-difference asymmetry 
  (as denoted in the panels) as functions of \(x\), \(z\),  or
  \(\pperp\). The systematic uncertainty is given as a band at the bottom of
  each panel. In addition there is a 7.3\%  scale uncertainty from the 
  target-polarization measurement. 
}
\end{figure}


Based on a \pythia \ Monte Carlo simulation~\cite{PYTHIA6} tuned to \hermes\ data,
the fraction of charged pions (kaons) stemming from the decay of exclusive 
vector-meson channels was estimated to be 
about  6--7\%  (2--3\%).
Among the contributions of all the vector mesons  to the pion samples, that of the 
\(\rho^0\) is dominant. A different observable, 
for which the contributions from exclusive \(\rho^0\) mesons cancels, 
is the {\em pion-difference asymmetry}
\begin{equation} \label{eq:pion-yield-difference}
A_{UT}^{\pi^+-\pi^-} \!(\phi, \phi_S)
\equiv 
 \frac{1}{ \mid \!\! \VSt \!\! \mid} 
 \frac{(\sigma_{{U}\uparrow}^{\pi^+}\!-\!\sigma_{{U}\uparrow}^{\pi^-}) - (\sigma_{{U}\downarrow}^{\pi^+} \!-\! \sigma_{{U}\downarrow}^{\pi^-})}{(\sigma_{{U}\uparrow}^{\pi^+}\!-\!\sigma_{{U}\uparrow}^{\pi^-}) + (\sigma_{{U}\downarrow}^{\pi^+}\!-\! \sigma_{{U}\downarrow}^{\pi^-})},
\end{equation}
the SSA in the difference
in the \(\pi^+\) and \(\pi^-\) 
cross sections for opposite
target-spin states \(\uparrow,\downarrow\).
In addition, this asymmetry helps to isolate the valence-quark Sivers functions: 
under some assumptions, such as charge-conjugation and isospin symmetry among
pion fragmentation functions, one can deduce from Eq.~\eqref{eq:QPM-sivers}
that this SSA stems mainly from the difference \((\siversf{, d_{v}}  - 4 \siversf{, u_{v}})\)
in the Sivers functions for valence down and up quarks.

The resulting Sivers amplitudes for pions, charged kaons, and
for the pion-difference asymmetry are shown in Fig.~\ref{fig:main-results-all} 
as functions of \(x\), \(z\), or \(P_{h\perp}\). 
They are positive and increase with increasing \(z\), 
except for \(\pi^-\), for which they are consistent with zero.
In the case of \(\pi^+\),  \(K^+\), and the pion-difference asymmetry,
the data suggest a saturation of the amplitudes for \(\pperp \gtrsim 0.4\)~GeV 
and are consistent with the predicted linear decrease in the limit of \(\pperp\) going to zero.

\begin{figure}
\centering
\includegraphics[scale=0.28]{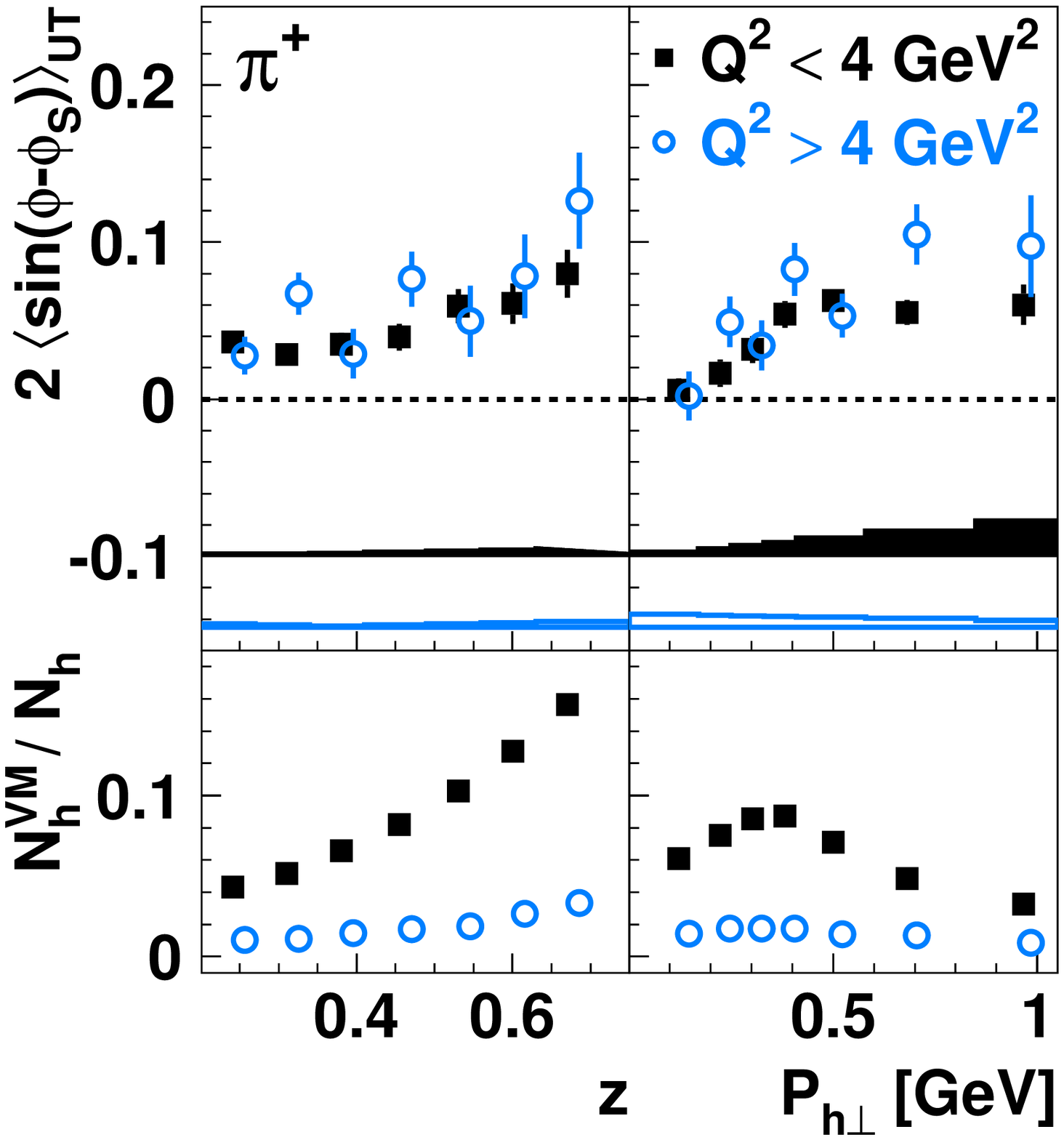}
\includegraphics[scale=0.28]{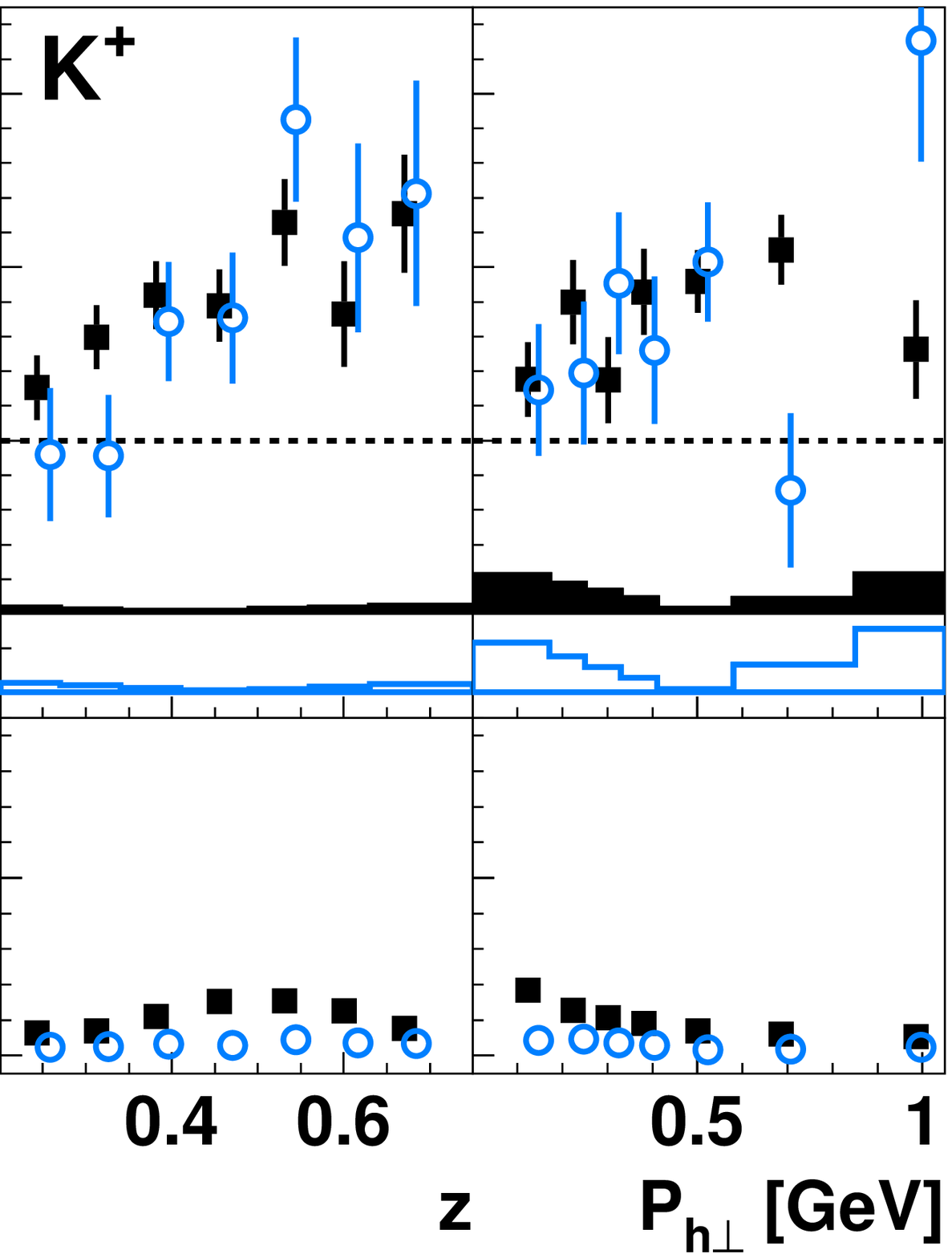}
\caption{\label{fig:Q2studyVM} Sivers amplitudes for \(\pi^+\) 
  (left) and \(K^+\) (right) as functions of \(z\) or \(\pperp\),
   compared for two different ranges in $Q^2$  (high-\(Q^2\) points are 
  slightly shifted horizontally). 
  The corresponding fraction of pions and kaons 
  stemming from exclusive vector mesons, extracted 
  from a Monte Carlo simulation, is provided in
  the bottom panels.
}
\end{figure}

\begin{figure}
\centering
\includegraphics[scale=0.28]{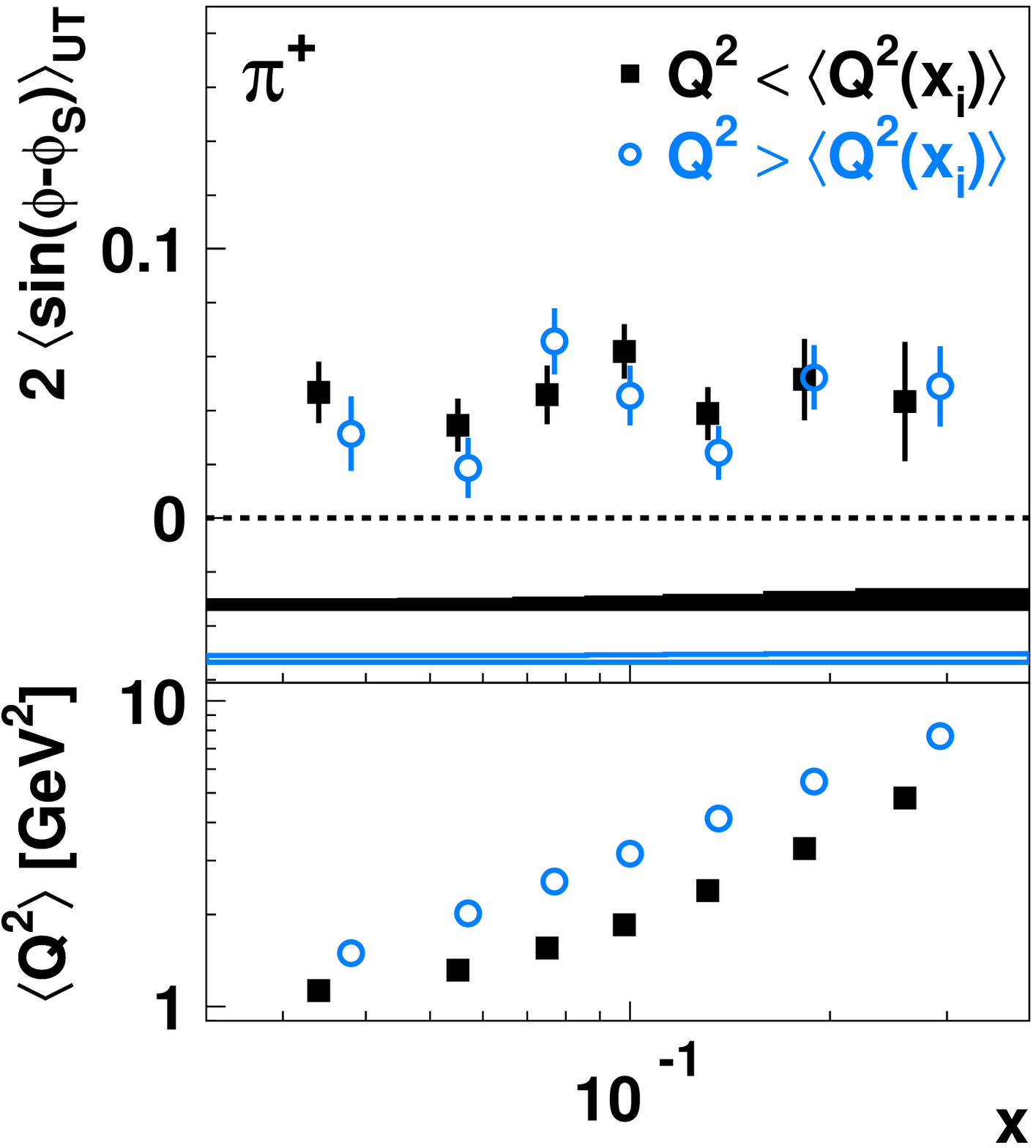}
\includegraphics[scale=0.28]{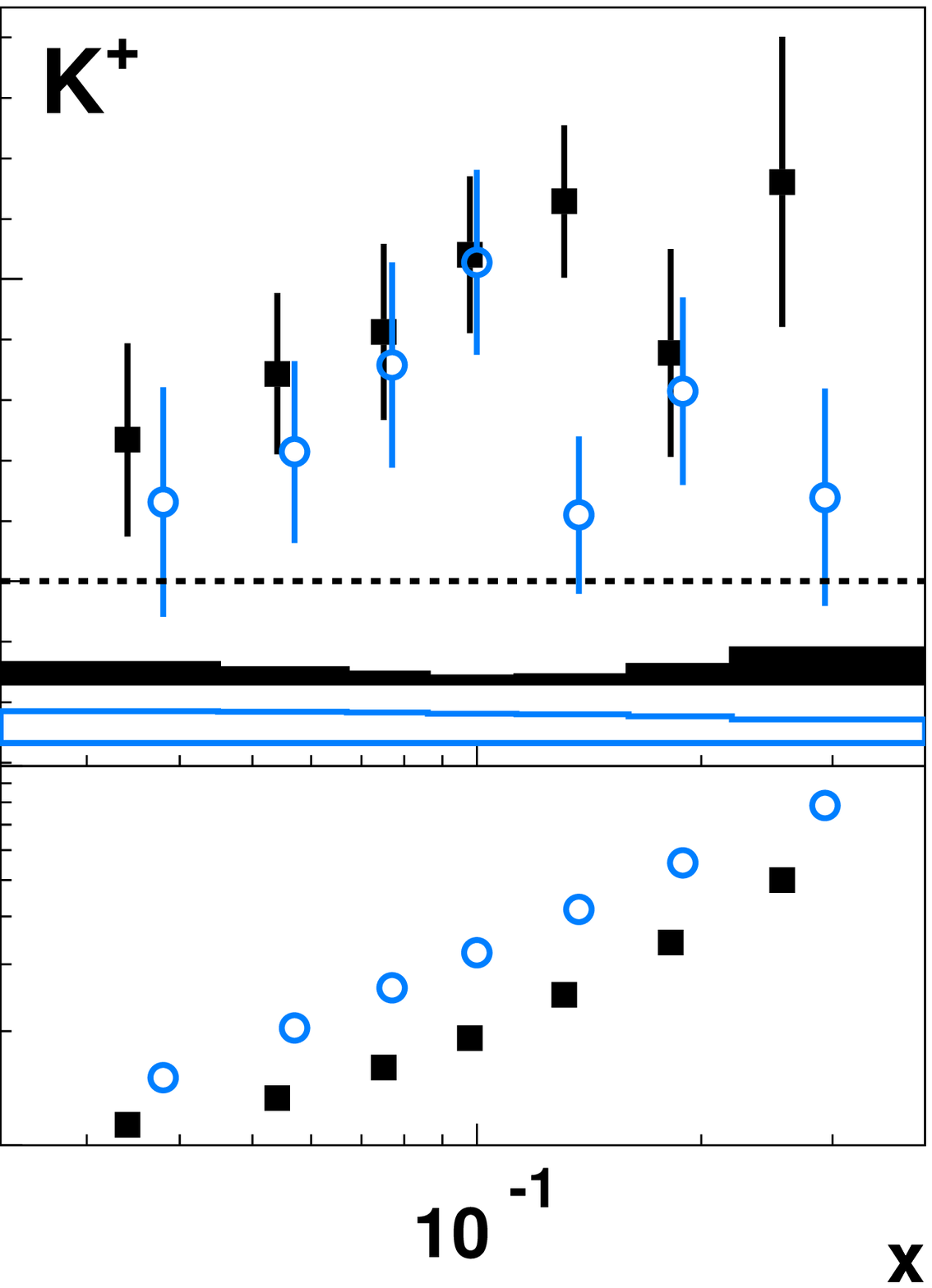}
\caption{\label{fig:Q2studyTwist} 
  Sivers amplitudes for \(\pi^+\)  (left) and \(K^+\) (right)
  as functions of \(x\). The \(Q^2\) range for each bin was 
  divided into the two regions above and below 
  \(\langle Q^2(x_i)\rangle\) of that bin. In the bottom 
  the average \( Q^2\) values  are given for the two \(Q^2\) ranges.
}  
\end{figure}

In order to further examine the influence of exclusive vector-meson decay and 
other possible \(\frac{1}{Q^2}\)-suppressed
contributions, several studies were performed. 
Raising the lower limit of \(Q^2\) to 4~GeV\(^2\) eliminates a large
part of the vector-meson contribution.
Because of strong correlations between \(x\) and \(Q^2\) in the data, 
this is presented only for the \(z\) and \(\pperp\) dependences. 
No influence of the vector-meson fraction on the asymmetries is visible as shown in
Fig.~\ref{fig:Q2studyVM}.
For the \(x\) dependence shown in Fig.~\ref{fig:Q2studyTwist},
each bin was divided into two \(Q^2\) regions below and above the corresponding  
average \(Q^2\)  (\(\langle Q^2(x_i)\rangle\)) for that \(x\) bin.
While the averages of the kinematics integrated over in those \(x\) bins do not differ significantly,
the \(\langle Q^2 \rangle\) values for the two \(Q^2\) ranges change by a factor of about 1.7. 
The asymmetries do not change by as much
as would have been expected for a sizable  \(\frac{1}{Q^2}\)-suppressed contribution, 
e.g., the one from longitudinal photons to the spin-(in)dependent cross section.
However, while the \(\pi^+\) asymmetries for the two \(Q^2\) regions are fully consistent, 
there is a hint of systematically smaller \(K^+\) asymmetries in the large-\(Q^2\) region.

%
%

\begin{figure}
\centering
\includegraphics[scale=0.485]{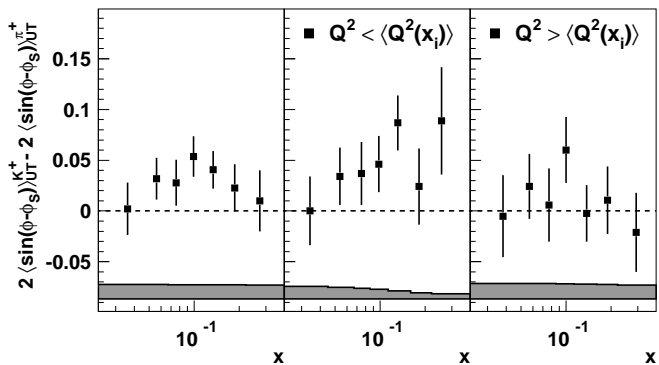}
\caption{\label{fig:piKstudy} Difference of Sivers amplitudes for \(K^+\) and \(\pi^+\)
  as functions of \(x\) for all \(Q^2\) (left), and separated into "low-" and 
  "high-\(Q^2\)" regions as done for Fig.~\ref{fig:Q2studyTwist}.}
\end{figure}

An interesting facet of the data is the difference in the \(\pi^+\)  and \(K^+\) amplitudes 
shown in Fig.~\ref{fig:piKstudy}.
On the basis of \(u\)-quark dominance, i.e., the dominant contribution to \(\pi^+\) and \(K^+\)
production from scattering off \(u\)-quarks, 
one might naively expect that the \(\pi^+\) and \(K^+\)  amplitudes should be similar.
The difference in the \(\pi^+\) and \(K^+\) amplitudes may thus point
to a significant role of other quark flavors, e.g., sea  quarks. 
Strictly speaking, even in the case of scattering solely off \(u\)-quarks, 
the fragmentation function \(D_1\), contained in both the numerator and denominator in 
Eq.~\eqref{eq:QPM-sivers}, does not cancel in general as it appears in convolution integrals.
This can lead not only to additional \(z\)-dependences, but also to a difference in size of the 
Sivers amplitude for \(\pi^+\) and \(K^+\).
Higher-twist effects in kaon production might also contribute to the difference observed:
in the low-\(Q^2\) region, 
where higher-twist should be more pronounced, the  \(\pi^+\) and \(K^+\) 
amplitudes disagree at the confidence level of
at least 90\%, based on a Student's \(t\)-test,
while being statistically consistent in the high-\(Q^2\) region.

As scattering off $u$-quarks dominates in these data, the positive Sivers amplitudes for \(\pi^+\) and \(K^\pm\)
suggest a large and negative Sivers function for $u$-quarks. This is supported by the positive 
amplitudes of the difference asymmetry, which is dominated by the contribution from valence \(u\)-quarks.
The vanishing amplitudes for \(\pi^-\) require cancelation effects, e.g., from a \(d\)-quark Sivers function 
opposite in sign to the \(u\)-quark Sivers function. 
In combination with deuteron data from the \compass\ collaboration~\cite{Ageev:2006da}, 
a large positive $d$-quark Sivers function can be deduced~\cite{Vogelsang:2005cs}. 
These fits have yet to be updated with the final results presented here, as well as with preliminary 
proton data from \compass~\cite{Levorato:2008tv}.

%
%

In summary, non-zero Sivers amplitudes in semi-inclusive DIS were measured 
for production of \(\pi^+\), \(\pi^0\), 
and \(K^\pm\), as well as for the pion-difference asymmetry. 
They can be explained by the non-vanishing naive-T-odd, 
transverse-momentum-dependent
Sivers distribution function. This function
also plays an important role in transverse single-spin 
asymmetries in \(pp\) collisions, and is linked to orbital angular momentum 
of quarks inside the nucleon. Although no
quantitative  conclusion about their orbital angular momentum can be inferred, 
the Sivers function provides important constraints on 
the nucleon wave function and thus indirectly on
the total quark orbital angular momentum.
For instance, in the approach of Ref.~\cite{Burkardt:2003yg}, 
the measured positive Sivers asymmetries for \(\pi^+\)
and \(K^+\) mesons correspond to a positive contribution of 
\(u\)-quarks to the orbital angular momentum,
under the assumption that the production of \(\pi^+\)
and \(K^+\) mesons is dominated by scattering off \(u\)-quarks.

%
%

\begin{acknowledgments} 
We gratefully acknowledge the \desy \ management for its support, the staff
at \desy \ and the collaborating institutions for their significant effort,
and our national funding agencies and the EU RII3-CT-2004-506078 program
for financial support.
\end{acknowledgments}

%
%


\end{document}